\documentclass{article}                     

\title{Editorial note to
``Large number coincidences\\ and the anthropic principle in
cosmology''} \author{George Ellis, University of Cape Town}

\begin{document}
\maketitle
\begin{abstract} This is an editorial note to accompany
reprinting as a Golden Oldie in the \emph{Journal of General
Relativity and Gravitation}\footnote{See
http://www.mth.uct.ac.za/~cwh/goldies.html.} of the famous paper
by Brandon Carter on the anthropic principle in cosmology
\cite{Car74}. This paper was presented at IAU Symposium No. 63,
entitled {\it Confrontation of cosmological theories with
observational data}, in 1973.\footnote{That republication will
take place later this year.}
\end{abstract}

The anthropic principle is one of the most controversial proposals
in cosmology. It relates to why the universe is of such a nature
as to allow the existence of life. This inevitably engages with
the foundations of cosmology, and has philosophical as well as
technical aspects. The literature on the topic is vast -- the
Carter paper reprinted here \cite{Car74} has 226 listed citations,
and the Barrow and Tipler book \cite{BarTip86} has 1740. Obviously
I can only pick up on a few of these papers in this brief review
of the impact of the paper.\\

While there were numerous previous philosophical treatises on the
topic, stretching back to speculations about the origin of the
universe in ancient times (see \cite {BarTip86} for a magisterial
survey), scientific proposals are more recent. A well known one was
biologist Alfred Russell Wallace, who wrote in 1904: ``Such a vast
and complex universe as that which we know exists around us, may
have been absolutely required ... in order to produce a world that
should be precisely adapted in every detail for the orderly
development of life culminating in man'' \cite{Wall04}. But that was
before modern cosmology was established: the idea of the expanding
and evolving universe was yet to come.

Following on Weyl's speculations on the fundamental physical constants in 1919,
and Eddington's work from 1923 on (summarised in his book \cite{Edd48}) hoping
to explain the numerical values of these constants,\footnote{See pp.224 -- 231
of \cite{BarTip86} for detailed references.} Dirac's ruminations on the large
number coincidences between these constants \cite{Dir37} seem to have set the
ball rolling. He suggested \cite{Dir38} that these coincidences in the evolving
universe had to be explained by a time varying gravitational constant, else they
would only be true at one cosmic epoch.

Robert Dicke \cite{Dic61} noted in 1961 that Dirac's relation
could be a selection effect related to the times when observers
could exist, this in turn being related to the values of the
fundamental constants of physics which determine the lifetime of
main sequence stars. Dicke's paper makes an explicit anthropic
statement in the cosmological context, based on an analysis of the
parameters affecting stellar lifetimes: ``$T$ [the age of the
universe] is not a random choice from a wide range of possible
choices but is limited by the criteria for the possible existence
of physicists''.\\

Brandon Carter, then a postdoc at Cambridge University, mused on
these topics, being inspired to do so first by the discussion of
Dirac's work in Bondi's classic book \emph{Cosmology} \cite{Bon60},
and then by Dicke's paper \cite{Dic61}. In 1967 he wrote a preprint
\cite{Car67} systematically investigating the causal links between
these constants and both astronomical and physical phenomena,
focusing first on the mass $m_N$ of the nucleon and its relation to
the electric charge $e$ and the mass of the electron $m_e$, and then
including three more sufficient to determine ordinary everyday
physics (the strong interaction coupling constant $g_S$, the pion
mass $m_\pi$, and the mass difference $\Delta_N$ between the neutron
and the proton) and three relating to more exotic processes (the
muon mass $m_\mu$, the electromagnetic mass splitting $\Delta_\pi$
of the pion, and the weak coupling constant $g_W$). He commented on
some remarkable relations between these parameters (equations (6) --
(8) of \cite{Car67}). He calculated many astrophysical consequences
of these numbers, but did not make any explicit anthropic link. A
second preprint in 1968, that was less widely circulated, made the
anthropic link to their values in the context of an ensemble of
universes.\footnote{This preprint is referred to in the Collins and
Hawking paper (\cite{ColHaw73}, p.319).}

This then opened the way for his conference talks on 21 February,
1970 at a Clifford Centenary meeting in Princeton,\footnote{
Carter, B. ``Large Numbers in Astrophysics and Cosmology''. Paper
presented at Clifford Centennial Meet., Princeton (1970);
unpublished. Referred to in \cite{CarRee79}.} and at an IAU
Symposium honouring Copernicus's 500th birthday held in Cracow
from September 10 to 17, 1973 (both talks being at the invitation
of John Wheeler). The latter talk was published in the symposium
proceedings in 1974 \cite{Car74}, which are reprinting as a Golden
Oldie. This definitively opened up the topic of the anthropic
principle as a subject for scientific discussion in the context of
present day cosmological theory. It was contemporaneous with a
paper by Collins and Hawking, ``Why is the universe isotropic?''
\cite{ColHaw73}, that raised the same issue in the specific
context of Bianchi cosmologies. The authors were all at the same
institution, so it was the subject of discussion between them at
that time; Carter's work is referenced in the Collins and Hawking
paper, which in turn is referenced in Carter's 1974
paper.\\

The Collins and Hawking paper was a specific application to the case
of Bianchi cosmologies, showing that only some Bianchi classes would
be suitable for life to develop. It states ``One cannot explain the
isotropy of the universe without postulating special initial
conditions ... we have to face the awkward question, why is the
universe isotropic?'' Like Carter they consider an ensemble of
universes. After explaining that only in Bianchi universes that tend
to become isotropic are galaxies likely to form, the paper states
\begin{quote}
\textbf{SAP (Collins and Hawking 1973)}: ``Since it would seem that
existence of galaxies is a necessary condition for the development
of intelligent life, the answer to the question `why is the universe
isotropic?' is `because we are here' ''.
\end{quote}
This statement appears to invert the expected direction of
causality, and so led to some suspicion about the anthropic
principle in many circles; however it in fact was to be seen as an
explanatory proposal in the context of a multiverse (which was
mentioned earlier in the paper).\\

The physics based approach by Carter in his paper \cite{Car74}
opened up much interest in the Anthropic Principle as a scientific
method of explanation. In addition to the mass of the proton $m_p$,
the mass of the electron $m_e$ and the electric charge $e$, like
Dicke he introduced the Hubble expansion rate $H$ (closely related
to the age of the universe $\simeq H^{-1}$), and related these to
other cosmological parameters and conditions required for the
existence of observers. He characterised the \emph{Weak Anthropic
Principle} (used by Dicke in \cite{Dic61}) as follows:
\begin{quote}
\textbf{WAP (Carter 1974)}: ``We must be prepared to take account of
the fact that our location in the universe is necessarily privileged
to the extent of being compatible with our existence as observers.''
\end{quote}
Thus it is seen as an observational selection effect in a universe
that is biofriendly; it relates to where and when we can exist in
such a universe (``location'' refers to time as well as space), and
is to be seen as a contrast to the Copernican principle (all places
are to be regarded as equivalent). He then proposed the \emph{Strong
Anthropic Principle}, relating to severe restrictions on the
fundamental parameters of the universe itself if life is to exist:
\begin{quote}
\textbf{SAP (Carter 1974)}: ``The universe (and hence the
fundamental parameters on which it depends) must be such as to admit
the creation of observers within it at some stage''.
\end{quote}
This sounds like the Collins and Hawking statement. But this was
intended as a \emph{prediction}, not an explanation; it took the
existence of observers for granted, and made deductions on that
basis. However he then introduced the idea that an ensemble of
universes was necessary in order that the conditions for life to
exist could be explained in a probabilistic way:
\begin{quote}
\textbf{SAP+ (Carter 1974)} ``It is of course philosophically possible -- as a
last resort, when no stronger physical argument is available -- to promote a
\emph{prediction} based on the strong anthropic principle to the status of an
\emph{explanation} by thinking in terms of a `world ensemble'.''
\end{quote}
This is the most common interpretation of the principle at the
present time.\\

The topic was developed further in an important paper by Carr and
Rees \cite{CarRee79}, broadening the set of relationships between
the constants considered. The abstract of that paper states, ``The
basic features of galaxies, stars, planets and the everyday world
are essentially determined by a few microphysical constants and by
the effects of gravitation. Many interrelations between different
scales that at first sight seem surprising are straightforward
consequences of simple physical arguments. But several aspects of
our Universe -- some of which seem to be prerequisites for the
evolution of any form of life -- depend rather delicately on
apparent 'coincidences' among the physical constants."  A similar
discussion was given later by Davies \cite{Dav82}. This suggests
that there is something needing explaining, unless these
bio-friendly relations were just coincidences -- but scientific
progress often lies in showing that what at first seem like
coincidences are in fact necessary relations.\\

The topic was then picked up in a Royal Society discussion meeting
on the constants of nature organised in 1983 by W H McCrea, where
Carter stated the strong principle as follows:
\begin{quote}
\textbf{SAP (Carter 1983)}: ``Our mere existence as intelligent
observers imposes restrictions not merely on our situation, but even
on the general properties of the universe including the values of
the fundamental parameters''. \cite{Car83}
\end{quote}
This version again is not an explanation, but a deduction from the
fact of our existence. Carter explained that this did not imply
that life was probable; he predicted that the occurrence of
observers would be rare, even on environmentally favorable planets
such as ours.\\

The topic was then presented in an encyclopaedic way in the book by Barrow and
Tipler \cite{BarTip86}, covering old design arguments, modern teleology, various
Anthropic principles, and their relation to quantum mechanics and biochemistry.
This book became a major reference in the area and made it widely known. It
interpreted the Weak Anthropic Principle as a selection effect: ``any
cosmological observations made by astronomers are biassed by an all-embracing
selection effect: our own existence'' (\cite{BarTip86}, pp.15 -- 16), but this
selection effect is interpreted in a multiverse setting (p.19). Their version of
the SAP (p.21) is essentially the same as Carter's 1974 version. They introduced
two further versions: Wheeler's controversial \emph{Participatory Anthropic
Principle} (PAP), based on his interpretation of quantum theory: ``Observers are
necessary to bring the universe into being''; and their own \emph{Final
Anthropic Principle}:
\begin{quote}
\textbf{FAP (Barrow and Tippler 1986)}: ``Intelligent
information-processing must come into existence in the Universe,
and, once it comes into existence, it will never die out.''
\end{quote}
These two inclusions somewhat discredited an otherwise very fine
volume.  In his famous review of the book \cite{Gar86}, Martin
Gardner ridiculed the FAP by quoting the last two sentences of the
book, which he characterized as representing the Completely
Ridiculous Anthropic Principle (CRAP). These versions (PAP and
FAP) have now faded into obscurity.\\

The publication of this book opened the floodgates for more
philosophically based discussions as well as popular presentations
(e.g. \cite{Ear87,Les89,Teg98,Ree00,Hog00,DemLam94,Bos02}). An
excellent survey article was written by Balashov \cite{Bal91}, with
extensive references up to 1991. The more the issue of fine tuning
was investigated, the more pressing it became. Rees for example
\cite{Ree00} argued that anthropic reasoning is needed to explain
the fine-tuned values of six constants required for life to exist:
\begin{itemize}
  \item $ N \simeq 10^{36}$: the strength of the electrical forces that hold
electrons and protons together to form atoms, divided by the force
of gravity between them;
  \item $\epsilon \simeq 0.007$: defining how strongly atomic nuclei bind
together;
  \item $\Omega_m \simeq 0.3$: the actual density of the universe expressed
in terms of the critical density required for it to eventually
recollapse in the future, if there is no cosmological constant;
  \item $\Omega_\Lambda \simeq 0.7$: The energy density of the
cosmological constant, similarly expressed as a dimensionless ratio;
  \item $Q \simeq 10^{-5}$: The fractional density of the
fluctuations in the early universe that were seeds of structure
formation via gravitational attraction;
  \item $D = 3$: the number of spatial dimensions in our world (at
macroscopic scales).
\end{itemize}
The constants $N$, $\epsilon$ and $D$ govern the fundamental
interactions of physics; $\Omega_m$, $\Omega_\Lambda$ and $Q$
govern the expansion of the universe and the growth of structures
in it. These are not the only parameters that have to be fine
tuned, for example the proton-neutron mass difference is not
listed here. The interactions between physical constants are
complex, and indeed it is not clear which should be thought of as
fundamental (see \cite{Hog00,Bar02,Uza03,Uza11}); for example
these parameters are not precisely the same as those discussed by
Carter \cite{Car67,Car74}, although they are related.\\

Although the evidence for fine tuning was accumulating, the
anthropic principle project was initially not widely accepted: it
was regarded as a philosophical oddity in cosmology, not really
related to physics. It became much more widely accepted as a result
of three developments.

Firstly, there was a growing appreciation of the major problem of
the magnitude of the cosmological constant, when considered as the
vacuum energy of quantum fields. Simple quantum field theory
calculation suggest it should be many orders of magnitude larger
then the measured value \cite{Wei89}. This is a major problem for
theoretical physics, because this prediction arises from a plausible
combination of two of our most successful theories, namely general
relativity and quantum field theory. This has crucial implications
for astrophysics because a value much larger than observed would
prevent galaxy formation, and hence would result in a lifeless
universe. A multiverse solution seemed almost the only way out
\cite{Wei87,Wei89}.

Secondly, the proposal of chaotic inflation \cite{Lin86} arose as a
variant of Guth's inflationary universe proposal, and this seemed to
provide a natural setting for existence of many disconnected
expanding universe domains that could have different effective
physical constants in each of them. This could reasonably be
considered a physical mechanism for realization of a `multiverse',
as originally envisaged by Carter \cite{Car74} and Collins and
Hawking \cite{ColHaw73}.

Thirdly, the proposal of the landscape of string theory arose as a
physics setting within which variation of effective constants might
naturally arise \cite{Sus03,Sus05}. The huge number of string vacua
posited in that theory provided a setting whereby the effective
constants of physics could be different in different domains such as
occur in chaotic inflation.

Together these ideas gave a coherent explanation of an otherwise
troubling issue: why does the cosmological constant have a small
value that allows life to exist? Once one has this theoretical
setting, one can apply the same explanation to any other of the
`fundamental' physical constants that need fine tuning in order
that life can exist, such as those mentioned by Rees \cite{Ree00}:
they too can arise from a deeper theory (M-Theory) that allows
each of them to have values that vary with position in a
multiverse. One can even suggest that in this way, the anthropic
principle is indispensible for making sense of the hypothesized
landscape of superstring theory (Brandon Carter, private
communication). However one should note that the basic mechanism
for eternal inflation does not of itself cause physics to be
different in each domain in a multiverse. This supposed variation
is not a necessary conclusion arising from the straightforward
conjunction of the physics of chaotic inflation and M theory; but
it can be the outcome if one extends the theory by adding extra
assumptions.\\

Proponents of the SAP in its explanatory form include Andrei
Linde, Martin Rees, Stephen Weinberg, and Leonard Susskind, but
others such as Roger Penrose, David Gross, and Lee Smolin
disagree. Various authors (including Aguirre, Bostrom, Carr,
Carter, Davies, Dimopoulos, Donoghue, Hartle, Hogan, Kallosh,
Linde, Page, Rees, Smolin, Stoeger, Susskind, Vilenkin, Weinberg,
and Wilczek) summarize their views on anthropic proposals in
\cite{Car07}. A recent review of the principle by Carter is
\cite{Car04}.\\

Various criticisms have been made of the proposal; I summarize them
as follows:

First, Carter \cite{Car04} has regretted his use of the word
``anthropic'', because it conveys the impression that the principle
involves only humans, rather than general intelligent observers;
others have commented that the term is misleading because the usual
calculations have nothing directly to do with humans, but refer only
to complex atoms or stars and planets. However these remarks do not
detract from the fact that the principle as commonly used relates to
real restrictions that will plausibly apply to existence of any
materially-based form of intelligent life, and hence is indeed an
interesting topic of discussion. Nevertheless if one is to seriously
contemplate the relation of physics to the origin of life, one does
need to relate the theory to biological realities
\cite{Car83,Car04}.

Second, there has been criticism that the word ``principle'' is not
justified in describing what are in fact just selection effects, or
indeed perhaps tautologies. However selection effects play a crucial
role in science, and are far from trivial; and in particular,
examining the consequences of changes in the fundamental constants
for existence of life is a useful and complex task. It certainly
does succeed in showing that the physical conditions required for
the existence of life are very special relative to what might
conceivably have been \cite{Ell06}.

Third, the explanatory form of SAP has been criticized by many
physicists as simply giving up on physical explanation. Penrose
states it thus:\footnote{This and other useful quotes are given in
an extensive article on the anthropic principle in Wikipedia.} ``It
tends to be invoked by theorists whenever they do not have a good
enough theory to explain the observed facts.'' \cite{Pen89}. The
force of this argument lies in the fact that a multiverse proposal
can be used to explain virtually anything, so it has almost no
predictive power. I return to this issue in a moment.

Fourthly, the results obtained in this way are completely dependent
on the measure used. Different (equally plausible) ways of assigning
probabilities to candidate universes lead to totally different
anthropic predictions. For example, Trotta and Starkman
\cite{TroSta06} present an explicit example based on the total
number of possible observations observers can carry out, and
conclude that in absence of a fundamental motivation for selecting
one weighting scheme over another, the anthropic principle cannot be
used to explain the value of $\Lambda$. The issue of the measure in
a multiverse is unresolved, and is the subject of an ongoing debate.
Related to this, anthropic bounds obtained on physical constants
depend on the specific choice of the theory. For example, Ref.
\cite{HarKriPer06} claims that you can have a life-supporting
universe (with carbon-12) without the weak interactions.

Finally, the physical existence or not of multiverse is basically
untestable \cite{Ell06}, but is an essential component in the
explanatory use of the SAP; so it is not clear if this can truly
be regarded as a scientific theory, rather than a philosophical
principle. The issue is debated in the articles in \cite{Car07}.\\

Given these criticisms, how does it fare as regards the acid test:
Can it make new predictions, or only retrospectively justify known
relationships? There are two cases where this kind of reasoning may
be claimed to have made new predictions, rather than post hoc
justifying relations that we already knew.

Firstly, a remarkable pioneering paper by Fred Hoyle made a crucial
observation regarding nuclear reactions required to create carbon-12
in helium-burning red giant stars \cite{Hoy53,Hoy54}. In order to
create the observed abundance of carbon, the triple-alpha process
required an excited resonant state of carbon-12 with an energy level
of 7.68 Mev (\cite{Hoy54}, p.134). The existence of the Hoyle state
was experimentally confirmed shortly after by Willie Fowler; new
data on the state is still being obtained today \cite{Cheetal10}.
Since carbon is crucial to the existence of life as we know it, and
indeed probably for existence of any physical living beings (because
there seems to be no alternative to polypeptide chains for creating
molecules of the required complexity), one can legitimately regard
existence of Hoyle's energy level as an anthropic requirement. Hoyle
himself did not make that link at the time of his discovery, but
seems to have done so later \cite{Kra}; one can claim that anthropic
requirements predict existence of this state. This is an instance of
use of SAP to make a \emph{prediction}, as originally proposed by
Carter in his 1974 paper. However it has since been related to
values of the fine structure constant \cite{Eksetal09}; this enables
proposal of an anthropic \emph{explanation} of the nature of the
Hoyle state.

Secondly, an intriguing line of argument by Weinberg and
collaborators \cite{MarShaWei97} is that, due to a `Principle of
Mediocrity' espoused by Vilenkin, $\Lambda$ is not likely to be
much smaller than the limit allowing galaxies to form. On this
basis they concluded that if $\Omega_\Lambda$ turned out to be
much less than $0.6$, anthropic reasoning could not explain why it
was so small. This was at a time when work by Efstathiou and
others suggested that $\Lambda$ was positive, but the definitive
supernova and CMB data putting the value at $\Omega_\Lambda \simeq
0.7$ was not yet available. This almost counts as an anthropic
prediction of something as yet unknown (in reviewing this argument
more recently \cite{Wei07}, Weinberg states that the observed
value is still a bit low relative to this anthropic prediction).
This is an example of use of SAP as an explanatory principle, but
depends on the additional assumption of the Principle of
Mediocrity -- an unverifiable philosophical principle.\\

Taken together, these two cases show that the anthropic principle
can potentially provide new predictions that can be experimentally
tested. It is thus not always simply philosophy or tautology.
However one must be cautious here: agreement with such predictions
is supporting evidence for the multiverse hypothesis as an
explanation, but does not amount to a proof that it is correct. If
hypothesis $M$ (existence of a multiverse) leads to prediction $L$
(some range of values for $\Lambda$), it is simple logic that
observation of $L$ does not necessarily imply validity of $M$
(this is only true if there is no other possible cause $M1$ that
also implies $L$). Furthermore, because this is only a
probabilistic claim based on a further philosophical assumption,
(\emph{not} $L$) does not imply (\emph{not} $M$): indeed no value
of the cosmological constant can either prove or disprove
existence of a multiverse, which is the crucial element in use of
the anthropic principle for explanatory purposes. Nevertheless the
Weinberg argument certainly makes it more plausible: it provides
legitimate supporting evidence.\\

The need for the strong version of the anthropic principle would
drop away if ever fundamental physics attained a foundational
theory uniquely predicting the correct values of all the
fundamental constants (which do of course allow life to exist)
\cite{Car04}. But this would not solve the anthropic mystery, as
some have claimed \cite{KanPerZyt00}, rather it would deepen it by
changing its locus to a more fundamental level. What possible
\emph{explanation} could one then give for why theoretical physics
principles (variational principles, specific symmetry groups, etc)
would \emph{necessarily} lead to the existence of life? It is not
clear how the issue could be tackled then, even in principle. We
are however apparently saved from this dilemma because our best
approach to developing a fundamental theory of all forces
(M-Theory) is presently heading in the opposite direction, by
predicting more and more possible quantum vacua and associated
effective theories \cite{Sus03,Sus05}. The needed uniqueness is
eluding us.\\

The anthropic principle continues to be the subject of much
controversy today, some regarding it as a crucial foundational
principle for cosmological argumentation, and others as giving up on
science. The debate continues. Its current application as an
explanatory principle is intimately tied in with the multiverse
hypothesis. As this is observationally untestable, one can argue
that it is a philosophical rather than strictly scientific proposal;
but as such, it is a very productive hypothesis, leading to many
interesting investigations. It deals with deep issues, and it will
not go away.

Despite what is often claimed by its proponents, the anthropic
principle used in this way does not solve issues of ultimate
explanation; it only postpones them. If this hypothesis is indeed
true, this simply leads to the next step in the argument: Why this
particular multiverse? Why any multiverse? How do the `laws'
governing the multiverse arise? Where do they
reside? And why are their governing constants what they are?\\

Carter's role in all of this was crucial: the paper \cite{Car04}
 initiated the vibrant discussion on this topic as a
scientific as opposed to purely philosophical enterprise. However as
is often the case,\footnote{Einstein's article ``The Foundation of
The General Theory of Relativity'' in \emph{Annalen der Physik}
\textbf{49} (1916) has only 246 citations recorded in Google Scholar
and 320 according to SPIRES.} after a relatively short while it was
replaced as a primary reference by other publications, and
particularly (as is evidenced by the citation figures quoted at the
start of this review essay) by the Barrow and Tipler book
\cite{BarTip86}. Carter himself moved on to discuss biological
issues and the probability of life in the universe, stating
\cite{Car83} ``The evidence suggests the evolutionary chain included
at least one but probably not more than two links which were highly
improbable (\emph{a priori}) in the time available'' . He regarded
this as an important result: in his 2004 review \cite{Car04}, he
states ``The example that seems to me most important was provided by
the prediction \cite{Car83} that the occurrence of anthropic
observers would be rare, even on environmentally favorable planets
such as ours''. Hence it is not enough to talk about carbon or
galaxies as being equivalent to the existence of observes (humans);
one needs to look at biological issues. One specific issue here is
whether the anthropic principle solves the crucial unsolved problem
of how the DNA code was created. It has been claimed by Eugene
Koonin that it can do so in the context
 of eternal inflation because
of the infinities proposed in that theory \cite{Koo07}, but this is
a controversial claim because those infinities are never realised in
a finite time \cite{EllSto09}.

Carter later formulated an anthropic interpretation of quantum
theory based on a microanthropic principle \cite{Car04a,Car07a}
relating to the foundations of quantum physics. This has not made
the same impact as the other versions. He is currently working on
issues to do with future and past of life on earth, considered in
the light of anthropic measures of hominid and other terrestrial
species
\cite{Car11}.\\

I thank Jean-Philippe Uzan and Jean-Pierre Lasota-Hirszowicz for
helpful comments on an earlier version of this comment.

\end{document}